\RequirePackage{fix-cm}
\documentclass[twocolumn,epjc3]{mysvjour3}  
\smartqed  
\RequirePackage{graphicx}
\RequirePackage{xcolor}
\RequirePackage{amssymb}
\RequirePackage{esint}
\newcommand{\eqref}[1]{(\ref{#1})}
\journalname{Eur. Phys. J. C}

\def\XXint#1#2#3{{\setbox0=\hbox{$#1{#2#3}{\int}$}
\vcenter{\hbox{$#2#3$}}\kern-.5\wd0}}

\def\TH{{\rm \Theta}}
\newcommand{\bqa}{\begin{eqnarray}}
\newcommand{\eqa}{\end{eqnarray}}
\newcommand{\nl}{\nonumber \\}

\newcommand{\s}{~\,}

\begin{document}

\title{Monte Carlo evaluation of divergent one-loop integrals without contour deformation
}
\author{Roberto Pittau\thanksref{e1,addr1}
}
\thankstext{e1}{e-mail: pittau@ugr.es}
\institute{Departamento de F\'isica Te\'orica y del Cosmos and CAFPE, Universidad de Granada, 18071 Granada, Spain\label{addr1}
}

\date{}
\date{Received: date / Accepted: date}

\maketitle

\begin{abstract}
  Reference \cite{Pittau:2021jbs} introduces a method for computing numerically four-dimensional multi-loop integrals without performing an explicit analytic contour deformation around threshold singularities. 
  In this paper, we extend such a technique to massless scalar one-loop integrals regularized in the framework of dimensional regularization. A two-loop example is also discussed.
\end{abstract}

\section{Introduction}
\label{sec:sec1}
In recent years a huge effort has been devoted to the problem of the computation of loop integrals. The reason is the need of accurate theoretical predictions able to cope with the ever-increasing precision of the data collected in particle physics experiments.

Two competing strategies have appeared. On the one hand, analytic methods based on systems of differential equations \cite{Kotikov:1991pm,Gehrmann:1999as,Henn:2013pwa}, whose solution are the wanted loop integrals, have shown their ability to cope with calculations involving a moderate number of physical scales \cite{Caola:2014iua,Gehrmann:2015bfy,Bonciani:2016qxi,Badger:2017jhb,Kudashkin:2017skd,Frellesvig:2019byn,Canko:2020ylt,Agarwal:2021vdh,Abreu:2021asb,Agarwal:2023suw,Abreu:2023rco}. On the other hand, techniques have been developed to deal with the problem in a fully numerical way \cite{Binoth:2003ak,Bierenbaum:2010cy,Runkel:2019yrs,Capatti:2020xjc,Liu:2022chg,Dubovyk:2022frj,Armadillo:2022ugh,Borinsky:2023jdv,Heinrich:2023til}.

The first obvious hurdle to overcome in both cases is the presence of infrared (IR) or ultraviolet (UV) divergences, that need to be properly regularized. This is usually done by using dimensional regularization  \cite{Bollini:1972ui,tHooft:1972tcz}, even if four-dimensional methods have started to be used as a viable alternative \cite{Pittau:2012zd,Donati:2013iya,Pittau:2013qla,Donati:2013voa,Page:2015zca,Gnendiger:2017pys,Page:2018ljf,TorresBobadilla:2020ekr,LTD:2024yrb}.

Regardless of the approach employed for regularizing infinities, the finite part of the calculation is plagued by the presence of the so-called threshold singularities. These are integrable singularities avoided by the $+i \epsilon$ propagator prescription.
Such structures are not a problem for analytic calculations, but must be properly addressed when using numerical techniques.

In a previous paper \cite{Pittau:2021jbs}, a method has been introduced, which permits an accurate fully numerical treatment of threshold singularities that can be easily implemented in Monte Carlo (MC) codes. The advantage of this technique is that, even if a non-zero $\epsilon$ must be kept, its influence on the result can be lowered close to the machine precision level, e.g. between $10^{-12}$ and $10^{-9}$ times the largest physical scale appearing in the problem.

The performance of this procedure has been studied in  \cite{Pittau:2021jbs} in the case of finite multi-loop Feynman integrals, or divergent ones regularized via four-dimensional methods. In this paper we extend for the first time this approach to integrals regularized within dimensional regularization, focusing our attention on scalar integrals that provide a complete basis for any one-loop calculation in massless theories \cite{Passarino:1978jh,Ossola:2006us}. Higher-loop integrals will be studied in detail elsewhere, although we envisage that the experience gathered at the one-loop level can be comfortably adapted to multi-loop environments.

The structure of the paper is as follows. In Sect.~\ref{sec:sec2} we briefly review the method. Section~\ref{sec:sec3} fixes our kinematics and conventions. Sections~\ref{sec:sec4},~\ref{sec:sec5} and~\ref{sec:sec6} are devoted to the study of the massless 2-, 3- and 4-point scalar one-loop integrals, respectively, and in Sect.~\ref{sec:sec7} we present a simple two-loop example.

Finally, it is important to mention that throughout the paper we distinguish between the $\epsilon$ and $\varepsilon$ symbols. The latter parameterizes the $n$-dimensional loop integration, $n= 4-2 \varepsilon$, while the former denotes the contour deformation around single-pole singularities.  All results presented in this paper are produced with $\epsilon= 10^{-9}$.

\section{The method}
\label{sec:sec2}
In this section we briefly recall the method of \cite{Pittau:2021jbs}.
Our aim is to flatten the singular behavior of a threshold singularity parameterized as
\bqa
\label{eq:I0}
I = \int_{-1}^1 dx\, \frac{F(x)}{x+i\epsilon},
\eqa
where the numerator function $F(x)$ is regular in $x=0$.
To achieve this we introduce a complex integration variable
$
z= \alpha + i \beta
$
related to $x$ by
$
x+i\epsilon= e^{i \pi (1-z)}.
$
The requirement that $x$ remains real fixes the path in the complex $z$ plane to be
\bqa
\label{eq:path}
\pi \beta= \ln \frac{\epsilon}{\sin[\pi (1-\alpha)]},
\eqa
so that
\bqa
x= \frac{\epsilon}{\tan[\pi (1-\alpha)]}.
\eqa
Using now
\bqa
\label{eq:par1}
dz= d\alpha \left(1+i \frac{d \beta}{d \alpha}\right)=
    d\alpha \left(1+i \frac{x}{\epsilon}\right)
\eqa
gives
\bqa
\label{eq:I1}
I = -\frac{i \pi}{g_\epsilon} \int_{{\epsilon}/{\pi}}^{1-{\epsilon}/{\pi}}\!\!\!\!d\alpha 
\left( 1+i\frac{x}{\epsilon} \right) F(x),~~~ g_\epsilon := 1-\frac{2\epsilon}{\pi}.
\eqa

Two comments are in order.
Firstly, the integrand of \eqref{eq:I1} is now regular in $x=0$ for arbitrarily small values of $\epsilon$. In fact, the $\epsilon$ dependence is moved to the boundaries of the integration region, $x=\pm 1$, which are reached exactly only when  $\epsilon \to 0$. However, $x = \pm 1$ are far away from the threshold singularity of \eqref{eq:I0}. That explains why the algorithm survives tiny numerical values of $\epsilon$.
Secondly, if $F(x)$ contains branch cuts in the $x$ complex  plane, the fact that $x$ always lies on the real axis ensures that the right Riemann sheet is automatically taken when $-1 \le x \le 1$. Thus, compared to methods based on contour deformation, one does not have to worry about choosing a path that avoids the pole at $x= -i \epsilon$ without crossing any cut of $F(x)$.

Equation \eqref{eq:I1} is optimal for integrating over $\alpha$. To flatten
the integral over $\beta$, the parameterization complementary to 
\eqref{eq:par1} is needed, namely
\bqa
\label{eq:par2}
dz= d\beta \left(\frac{d \alpha}{d \beta}+i\right).
\eqa
However, \eqref{eq:path} implies that $\alpha$ is a two-valued function of $\beta$. Therefore, it is necessary to divide \eqref{eq:I1} into two parts
\bqa
\label{eq:Ifb}
I  &=& -\frac{i \pi}{g_\epsilon}
\int_{{\epsilon}/{\pi}}^{1/2}\!\! d\alpha \nl
&&\times\left[
 \left(1-i\frac{y_\alpha}{\epsilon}\right)F(-y_\alpha)
+\left(1+i\frac{y_\alpha}{\epsilon}\right)F(y_\alpha)
\right],
\eqa
where $y_\alpha:= {\epsilon}/{\tan(\alpha \pi)}$.
Inserting \eqref{eq:par2} into \eqref{eq:Ifb} gives
\bqa
\label{eq:Ifc}
I &=& - \frac{i \pi}{g_\epsilon} \int_{\beta_-}^{\beta_+}
d\beta \nl
&& \times \left[
 \left( \frac{\epsilon}{-y_\beta} +i\right) F(-y_\beta)
-\left( \frac{\epsilon}{ y_\beta} +i \right)F( y_\beta)
\right],
\eqa
with
\bqa
y_\beta:= e^{\pi\beta}\sqrt{1-\left(\frac{\epsilon}{e^{\pi \beta}}\right)^2},\s\beta_-= \frac{1}{\pi}\ln \frac{\epsilon}{\sin \epsilon},\s\beta_+= \frac{\ln \epsilon}{\pi},  \nonumber
\eqa
that is optimized for the integration over $\beta$.
In practice, equations \eqref{eq:Ifb} and \eqref{eq:Ifc} can be merged together by means of a multichannel MC approach \cite{Kleiss:1994qy},~\footnote{Additional MC channels can be superimposed, if needed.} so that the complete $1/(x+i \epsilon)$ behavior of \eqref{eq:I0} is flattened.

The described algorithm is implemented in the code {\tt GLoop} \cite{gloop}.
The present version is able to deal with integrals of the type
\bqa
\int_{-\infty}^\infty \prod_{j=1}^m \left(\frac{d \sigma_j}{\sigma_j \pm i \epsilon} \right) F(\sigma_1,\sigma_2,\ldots,\sigma_m),
\eqa
with $m$ up to 4. The numerical results presented in this paper require $m= 3$ at most. 

\section{Kinematics and loop integration}
\label{sec:sec3}
For the purposes of this work it is sufficient to consider a $p_1+p_2 \to p_3+p_4$ massless kinematics given by
\bqa
\label{eq:vec}
\begin{tabular}{ll}
  $\!\!\!p^\alpha_1 = \frac{\sqrt{s}}{2} (1,1,0,0),$ &
  $\!\!\!\!p^\alpha_2 = \frac{\sqrt{s}}{2} (1,-1,0,0),$ \\
  $\!\!\!p^\alpha_3 = \frac{\sqrt{s}}{2} (1,\cos \theta_{13}, \sin \theta_{13},0),$ &
  $\!\!\!\!p^\alpha_4 = p^\alpha_1+p^\alpha_2-p^\alpha_3.$
\end{tabular} 
\eqa
In \eqref{eq:vec} $s=(p_1+p_2)^2$ and $\theta_{13}$ is the scattering angle defined by the relation
\bqa
t= (p_1-p_3)^2= -\frac{s}{2}(1-\cos\theta_{13}).
\eqa
The $n$-dimensional loop momentum is
\bqa
q^\alpha &=& (q_0,|\vec q|c_\theta,|\vec q|s_\theta c_\phi, \ldots),
\eqa
where
$c_\theta = \cos{\theta}$, $s_\theta = \sin{\theta}$, $c_\phi= \cos\phi$, with
$0 \le \theta \le \pi$ and $0 \le \phi \le 2 \pi$.

Rescaling $p_{1,2,3}$ and $q$ by $\sqrt{s}$ produces the following dimensionless vectors
\bqa
\label{eq:rescvec}
\begin{tabular}{ll}
  $\pi^\alpha_1 = \frac{1}{2} (1,1,0,0),$ &
  $\pi^\alpha_2 = \frac{1}{2} (1,-1,0,0),$ \\
  $\pi^\alpha_3 = \frac{1}{2} (1,c_{13},s_{13},0),$ &
  $\omega^\alpha_{n} = (\tau,\rho c_\theta,\rho s_\theta c_\phi, \ldots),$
\end{tabular}
\eqa
in which $\tau= {q_0}/{\sqrt{s}}$ and $\rho= {|\vec q|}/{\sqrt{s}}$.
Note that $\pi^\alpha_{1,2,3}$ span a $3$-dimensional space, so that in the one-loop integrands one can trade $\omega^\alpha_{n}$ for its $4$-dimensional projection defined as
\bqa
\omega^\alpha= (\tau,\rho c_\theta,\rho s_\theta c_\phi,\rho s_\theta s_\phi),
~~{\rm with}~~s_\phi= \sin \phi.
\eqa
In fact
$\omega^2_n= \omega^2$ and  $\omega_n \cdot \pi_i=  \omega \cdot \pi_i~~\forall i=1,2,3$.

Finally, the integration over the rescaled loop momentum can be parameterized as
\bqa
\label{eq:intnt}
\int\!d^n \omega_n  = \int_n \int_{-\infty}^{\infty}\! d\tau,
\eqa
where $\int_n$ is the $(n-1)$-dimensional integration volume. In terms of
$\rho$, $\theta$ and $\phi$ it reads
\bqa
\label{eq:intn0}
\int_n = 
\frac{2 \pi^{\frac{n-3}{2}}}{{\rm \Gamma}\left(\frac{n-3}{2}\right)}
&\times& \int_{-1}^{1} d c_\theta\, (1-c^2_\theta)^{\frac{n-4}{2}} \nl
&\times& \int_{-1}^{1} d c_\phi\, (1-c^2_\phi)^{\frac{n-5}{2}}
\int_0^\infty d \rho\, \rho^{n-2}.
\eqa
If the integrand is independent of $\phi$, the integration over $c_\phi$ can be carried out analytically by using
\bqa
\int_0^{\pi} (\sin \phi)^m\,d\phi= \sqrt{\pi}\, \frac{{\rm \Gamma} \left(\frac{m+1}{2}\right)}{{\rm \Gamma} \left(\frac{m+2}{2}\right)},
\eqa
that gives
\bqa
\label{eq:intn}
\int_n = 
\frac{2 \pi^{\frac{n-2}{2}}}{{\rm \Gamma}\left(\frac{n-2}{2}\right)}
\int_{-1}^{1} d c_\theta\, (1-c^2_\theta)^{\frac{n-4}{2}}
\int_0^\infty d \rho\, \rho^{n-2}.
\eqa
Likewise, integrating over $c_\theta$ produces
\bqa
\label{eq:intn1}
\int_n = \frac{2 \pi^{\frac{n-1}{2}}}{{\rm \Gamma}\left(\frac{n-1}{2}\right)}
\int_0^\infty d\rho\, \rho^{n-2}.
\eqa

\section{The UV divergent one-loop 2-point integral}
\label{sec:sec4}
\begin{figure}
\vskip -4.9cm
\hskip -5.2cm
\includegraphics[width=6.5in]{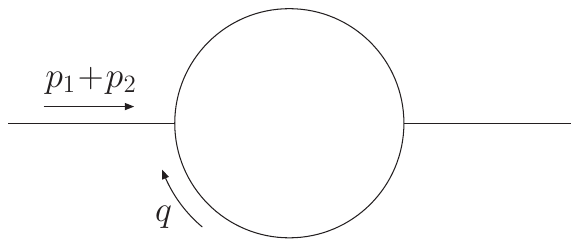}
\vskip -4.8cm
\caption{The scalar 2-point one-loop function of  \eqref{eq:BP1}.}
\label{fig:1}   
\end{figure}
As a first illustration  of our procedure we compute by MC the dimensionally regularized one-loop scalar integral of Fig.~\ref{fig:1},
\bqa
\label{eq:BP1}
B(s) = \mu^{4-n} \int d^n q \frac{1}{D_0 D_1},
\eqa
where
  $D_0= q^2+i \epsilon$ and
  $D_1= (q+p_1+p_2)^2+i \epsilon$.
Given that $B(s)$ diverges in four dimensions, our strategy is to split it into a finite part and a UV divergent piece,
\bqa
\label{eq:splitB}
B(s)= {B}_{\rm F}(s)+{B}_{\rm UV},
\eqa
in a way that the former can be computed numerically using the $4$-dimensional algorithm of Sect.~\ref{sec:sec2}, while the latter is evaluated analytically.

Rescaling loop and external momenta by $\sqrt{s}$ gives
\bqa
\label{eq:calB1}
B(s)= \left(\frac{s}{\mu^2} \right)^{-\varepsilon}\! {\cal B},~~{\rm where}~~
{\cal B}= \int_n \int_{-\infty}^\infty d \tau\, \frac{1}{d_0 d_1},
\eqa
with $d_0 = \tau^2-\rho^2+i \epsilon$ and $d_1 = (\tau+1)^2-\rho^2 +i \epsilon$.
The integrand of \eqref{eq:calB1} does not depend on $\theta$ and $\phi$.
Hence, $\int_n$ can be taken as in \eqref{eq:intn1}.
We now integrate over $\tau$ by using the residue theorem. The result is
\bqa
\label{eq:calB2}
    {\cal B}= \frac{i \pi}{2} \int_n\,\frac{1}{\rho} \frac{1}{\rho^2-1/4-i \epsilon}.
\eqa

To achieve the splitting  of \eqref{eq:splitB}, we subtract and add back 
an integrand with the same $\rho  \to \infty$  behavior of \eqref{eq:calB2}. Among the various possibilities we choose
$
\frac{1}{\rho(\rho^2+1/4)}.
$
This allows us to recast
$
{\cal B}= {\cal B}_{\rm F}+{\cal B}_{\rm UV}$, in which the finite piece reads
\bqa
\label{eq:calBfin1}
{\cal B}_{\rm F}= 2 i \pi^2
\int_0^\infty d \rho\,\rho \left[\frac{1}{\rho^2-1/4-i \epsilon}-
               \frac{1}{\rho^2+1/4}\right],
\eqa
while ${\cal B}_{\rm UV}$ is easily computed analytically,
\bqa
\label{eq:calBUV}
      {\cal B}_{\rm UV}= \frac{i \pi^{2-\varepsilon}}{{\rm \Gamma}(1 -\varepsilon)}
      \left(\frac{1}{\varepsilon} + 2\right) + {\cal O}(\varepsilon).
\eqa
Introducing the variable $\sigma= \rho^2-1/4$ allows one to rewrite
\eqref{eq:calBfin1} in a form suitable to be integrated with {\tt GLoop},
\bqa
\label{eq:calBfin2}
      {\cal B}_{\rm F} = \int_{-\infty}^{\infty} \frac{d \sigma}{\sigma-i \epsilon} F(\sigma),~~
      F(\sigma) =  i \pi^2\,\frac{\TH(\sigma+1/4)}{1 + 2 \sigma},
\eqa
where $\TH$ is the Heaviside step function. 

Our MC estimate with $10^{7}$ MC shots gives
\bqa
\label{eq:calBMC}
      {\cal B}_{\rm F}/(i \pi^2)=  3(9)\times 10^{-4}  +i\,3.1414(7),
\eqa
to be compared to the analytic value
\bqa
{\cal B}_{\rm F}/(i \pi^2) |_{\rm Analytic}= i\,\pi.
\eqa
The time to produce the result of \eqref{eq:calBMC} on a single 2.2 GHz processor is of about 1.6\,s.

Finally, the analytic continuation to $s < 0$ is achieved by replacing $s/\mu^2 \to
s/\mu^2 +i \epsilon$ in \eqref{eq:calB1} \cite{Ellis:2007qk}. This produces
\bqa
B(s)&=& \frac{i \pi^{2-\varepsilon}}{{\rm \Gamma}(1-\varepsilon)} \bigg[
   \frac{1}{\varepsilon} +2-L-i\pi+\frac{{\cal B}_{\rm F}}{i \pi^2} \bigg]
+ {\cal O}(\varepsilon),
\eqa
with
$L= \ln \big(-{s}/{\mu^2}- i \epsilon \big)$.

Lastly, it is noteworthy to mention that from a numerical standpoint, the treatment of UV divergences is remarkably similar in terms of dimensional regularization and FDR \cite{Pittau:2012zd,Donati:2013iya,Pittau:2013qla,Donati:2013voa,Page:2015zca,Page:2018ljf}. The sole distinction lies in their action towards the subtracted UV divergent part, which is computed analytically and added back in the former approach, whereas it is judiciously discarded in FDR \cite{Pittau:2021jbs}.
\section{The IR divergent one-loop 3-point integral}
\label{sec:sec5}
\begin{figure}
\vskip -4.2cm
\hskip -4.7cm
\includegraphics[width=6.5in]{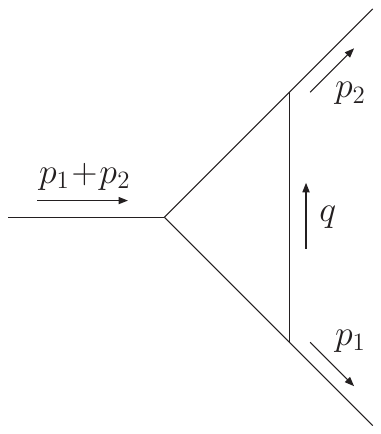}
\vskip -3.9cm
\caption{The scalar 3-point one-loop function of \eqref{eq:CP1}.}
\label{fig:2}   
\end{figure}
Here we study the dimensionally regularized triangle function of Fig.~\ref{fig:2} with two massless momenta
$p^2_{1,2}=0$ and massless denominators $D_0=q^2+i\epsilon$, $D_1=(q+p_1)^2+i\epsilon$ and $D_2=(q-p_2)^2+i\epsilon$,
\bqa
\label{eq:CP1}
 C(s) = \mu^{4-n} \int d^n q \frac{1}{D_0D_1D_2}.
\eqa
As in the previous section, we rescale all dimensionful quantities by $\sqrt{s}$. This produces
\bqa
\label{eq:CP2}
C(s) = \frac{1}{s}\left(\frac{s}{\mu^2} \right)^{-\varepsilon} {\cal C}. 
\eqa
The rescaled integral ${\cal C}$ reads 
\bqa
    {\cal C} = \int_n \int_{-\infty}^\infty d \tau\, \frac{1}{d_0d_1d_2}, 
\eqa
with $\int_n$ given in \eqref{eq:intn} and
\bqa
\label{eq:d012}
    d_0 &=& (\tau +\rho-i \epsilon) (\tau-\rho  +i \epsilon), \nl
    d_1 &=& (\tau +1/2+R-i \epsilon) (\tau+1/2-R+i \epsilon), \nl
    d_2 &=& (\tau -1/2+R-i \epsilon) (\tau-1/2-R+i \epsilon), \nl
     R  &=& \sqrt{1/4+\rho^2+\rho c_\theta}.
\eqa

${\cal C}$ is divergent in the soft and collinear limits, namely it develops $1/{\varepsilon}$ and $1/{\varepsilon^2}$ poles under the $n$-dimensional integration. To be able to perform the loop integration numerically, we first construct an approximation ${\cal C}_{\rm IR}$ whose integrand subtracts the infrared behavior in a local fashion. Then we reinsert the result of an analytic computation of ${\cal C}_{\rm IR}$. Schematically,
${\cal C}= {\cal C}_{\rm F}+{\cal C}_{\rm IR}$,
where
\bqa
\label{eq:eqsub}
{\cal C}_{\rm F}= \lim_{n \to 4}\left[{\cal C}-{\cal C}_{\rm IR}\right]
\eqa
is computed by MC.

Using the residue theorem to integrate over $\tau$ gives
\bqa
\label{eq:Cfun}
    {\cal C} &=& - \frac{i \pi}{2} \int_n \frac{1}{\rho^2 R }
    \left(
    \frac{1}{R-1/2-i \epsilon}-\frac{1}{R-1/2+\rho-i \epsilon} \right. \nl
      && \left. -\frac{1}{R+1/2}
         +\frac{1}{R+1/2+\rho}
    \right).
    \eqa
Only the first two terms of \eqref{eq:Cfun} are divergent when $\rho < 1/2$. An approximation sharing their IR behavior is constructed by expanding $R$ around $\rho(1+c_\theta)= 0$, that produces
\bqa
\label{eq:CCT}
    {\cal C}_{\rm IR} = i \pi \int_n \frac{\TH(1-2 \rho)}{\rho^3}
    \left(\frac{1}{1+c_\theta}-\frac{1}{c_\theta+2 \rho-i \epsilon}
    \right),
\eqa
which can be easily evaluated analytically
\bqa
    {\cal C}_{\rm IR}= \frac{i \pi^{2-\varepsilon}}{{\rm \Gamma}(1-\varepsilon)}
    \left(\frac{1}{\varepsilon^2}+ \frac{i \pi}{\varepsilon}
    -\frac{2 \pi^2 }{3} +i \pi \ln(4)
    \right) + {\cal O}(\varepsilon). \nonumber
\eqa

The integrand of \eqref{eq:CCT} can now be subtracted to \eqref{eq:Cfun} and the resulting integral produces the finite contribution of \eqref{eq:eqsub}. In four dimensions \eqref{eq:intn} gives
\bqa
\int_4= 4 \pi \int_0^\infty\!  d \rho\, \rho \int_{|R_-|}^{R_+}\! dR\, R,
\eqa
with $R_{\pm}= 1/2\pm \rho$, thus
\bqa
\label{eq:csub}
{\cal C}_{\rm F} &=& 
    -2 i \pi^2 \int_0^\infty\!  \frac{d \rho}{\rho}\,\int_{|R_-|}^{R_+}
    dR\, \bigg\{ \frac{1}{R-1/2-i \epsilon} \nl
    &&-\frac{{\TH(R_-)}}{R-\sqrt{R_+R_-}-i \epsilon}
    -\frac{\TH(-R_-)}{R-R_-}-\frac{1}{R+1/2} \nl
    && +\frac{1}{R+R_+}-\frac{\TH(R_-)}{R+\sqrt{R_+R_-}}
    +\frac{\TH(R_-)}{R+R_-} \bigg\},
\eqa
where the $-i \epsilon$ is kept only in denominators with threshold singularities.

Now we put \eqref{eq:csub} in a form suitable to be integrated with {\tt GLoop}.
The terms between curly brackets have
denominators of the form $R+r_i$, where the $r_{1 \div7}$ are independent of $R$. We then introduce two integration variables defined as follows
\bqa
\sigma_1= \rho,~\sigma_2&=& R+r_i~~\forall\, i= 1 \div 7,
\eqa
and rewrite
\bqa
\label{eq:csub1}
{\cal C}_{\rm F}  = 
\int_{-\infty}^{\infty} \prod_{j=1}^2 \left( \frac{d \sigma_j}{\sigma_j-i \epsilon}\right) F_{\cal C}(\sigma_1,\sigma_2).
\eqa
The numerator $F_{\cal C}(\sigma_1,\sigma_2)$ of \eqref{eq:csub1} is fully expressible in terms of Heaviside functions and is given in \ref{app:a}.

With $4 \times 10^8$ MC points our estimate is
\bqa
    {\cal C}_{\rm F}/(i \pi^2) = 1.644(4) -i\,4.356(1),
\eqa
to be compared to the analytic result
\bqa
\label{eq:CFana}
      {\cal C}_{\rm F}/(i \pi^2) |_{\rm Analytic} &=& \pi^2/6-i \pi \ln(4) \nl
      &=& 1.6449 -i\,4.3552.
\eqa
The time to produce $10^{6}$ MC shots on a single 2.2 GHz processor is of about 0.32\,s.

In our computation we have assumed $s > 0$. The analytic continuation to $s < 0$ is again obtained by replacing $s \to s+i \epsilon$ in \eqref{eq:CP2}. Expanding in $\varepsilon$ gives
\bqa
\label{eq:Cana}
C(s)&=& \frac{i \pi^{2-\varepsilon}}{{\rm \Gamma}(1-\varepsilon)}\frac{1}{s} \bigg[
  \frac{1}{\varepsilon^2}-\frac{L}{\varepsilon}+\frac{L^2}{2} -\frac{\pi^2 }{6} +i \pi \ln(4)\nl
   && + \frac{{\cal C}_{\rm F}}{i \pi^2} \bigg]
+ {\cal O}(\varepsilon),
\eqa
where
$L= \ln \big(-{s}/{\mu^2}- i \epsilon \big)$.

\section{The IR divergent one-loop 4-point integral}
\label{sec:sec6}
\begin{figure}
\vskip -4.4cm
\hskip -4.3cm
\includegraphics[width=6.5in]{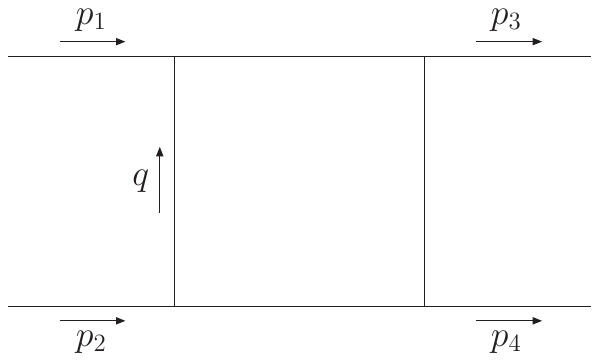}
\vskip -4.35cm
\caption{The scalar 4-point one-loop function of \eqref{eq:D1}.}
\label{fig:3}   
\end{figure}
In this section we consider the massless box diagram of Fig.~\ref{fig:3},
\bqa
\label{eq:D1}
&&D(s,t) = \mu^{4-n} \int d^nq \frac{1}{D_0 D_1 D_2 D_3}, \\ 
&&
\begin{tabular}{ll}
$D_0 = q^2+i\epsilon$, &  $D_1=(q+p_1)^2+i\epsilon$, \\
$D_2=(q-p_2)^2+i\epsilon$, & $D_3=(q+p_1-p_3)^2+i\epsilon$, \nonumber
\end{tabular}  
\eqa
where $p^2_{1,2,3,4}= 0$.
By rescaling all dimensionful quantities by $\sqrt{s}$ one arrives at 
\bqa
D(s,t) =\frac{1}{s^2} \left(\frac{s}{\mu^2}\right)^{-\varepsilon}  {\cal D}(x),
\eqa
where we have defined $x = -{t}/{s}$, so that in the physical region one has $0 \le x \le 1$. 
The rescaled 4-point integral reads
\bqa
\label{eq:rescd}
{\cal D}(x)= \int_n \int_{-\infty}^{\infty} d\tau\, \frac{1}{d_0d_1d_2d_3},
\eqa
with $\int_n$ given in \eqref{eq:intn0}.
The denominators $d_{0,1,2}$ are as in \eqref{eq:d012}. Furthermore,
$
d_3 = (\tau+S-i \epsilon) (\tau-S  +i \epsilon)
$ with
$S  = \sqrt{\rho^2+U-V\, c_\phi}$ and
\bqa
U = x (1 +2 \rho c_\theta),\quad V = 2 \rho \sqrt{x(1-x)} \sqrt{1 -c_\theta^2}.
\eqa

As before, to compute \eqref{eq:rescd} by MC, we first have to subtract a simpler function ${\cal D}_{\rm IR}(x)$ with the same local IR behavior of ${\cal D}(x)$. We choose 
\bqa
\label{eq:rescdsub}
    {\cal D}_{\rm IR}(x)&=& \int_n \int_{-\infty}^{\infty} d\tau\,
    \bigg[\frac{1}{d_0d_1d_3}+\frac{1}{d_0d_2d_3} 
         -\frac{1}{x}\,\frac{1}{d_0d_1d_2}\nl && -\frac{1}{x}\,\frac{1}{d_1d_2d_3}
      \bigg]-\frac{i\pi^2}{x}\left(\pi^2-\ln^2(x)\right), 
\eqa
that can be integrated analytically by means of \eqref{eq:CFana} and \eqref{eq:Cana},
\bqa
{\cal D}_{\rm IR}(x)= 
\frac{i \pi^{2-\varepsilon}}{{\rm \Gamma}(1-\varepsilon)} \frac{1}{x}
\left(-\frac{4}{\varepsilon^2}+ \frac{2}{\varepsilon}\left(\ln x -i \pi \right)
\right) + {\cal O}(\varepsilon). \nonumber
\eqa
The rationale behind \eqref{eq:rescdsub} is as follows. On the one hand, it is well known that the four 3-point integrals produce the same $1/\varepsilon$ and $1/\varepsilon^2$ poles of ${\cal D}(x)$ \cite{Dittmaier:2003bc}. On the other hand, the last term is chosen in such a way that it compensates their finite contribution. In this way
\bqa
\label{eq:DF}
{\cal D}_{\rm F}(x)= \lim_{n \to 4}\left[{\cal D}(x)-{\cal D}_{\rm IR}(x) \right]
\eqa
gives the finite part of ${\cal D}(x)$ directly.

Integrating over $\tau$ the IR finite combination of integrals appearing in \eqref{eq:DF} gives
\bqa
\label{eq:calD}
&&\int_4 \int_{-\infty}^{\infty} d\tau\, \bigg( \frac{1}{d_0d_1d_2d_3}
-\frac{1}{d_0d_1d_3}-\frac{1}{d_0d_2d_3}  \nl
&&~~ +\frac{1}{x}\,\frac{1}{d_0d_1d_2}+\frac{1}{x}\,\frac{1}{d_1d_2d_3}
\bigg) = \nl
&&i \pi \int_4 \frac{1}{\rho^2 R }
\bigg\{
          \frac{1}{R-1/2-i \epsilon}      \bigg( {\rm P}\bigg[\frac{1}{S^2-\rho^2}\bigg]-\frac{1}{x} \bigg) \nl
    &&~~ -\frac{1}{R-1/2+\rho-i \epsilon} \bigg( \big(1-2\rho\big) {\rm P}\bigg[\frac{1}{S^2-\rho^2}\bigg]-\frac{1}{x} \bigg) \nl
    &&~~ -\frac{1}{R+1/2}                 \bigg( {\rm P}\bigg[\frac{1}{S^2-\rho^2}\bigg]-\frac{1}{x} \bigg) \nl
    &&~~ +\frac{1}{R+1/2+\rho}            \bigg( \big(1+2\rho\big) {\rm P}\bigg[\frac{1}{S^2-\rho^2}\bigg]-\frac{1}{x} \bigg)
    \bigg\},
\eqa
where ${\rm P}$ denotes the Cauchy principal value,
\bqa
    {\rm P}\bigg[\frac{1}{S^2-\rho^2}\bigg] = \frac{1}{2}\bigg(\frac{1}{S^2-\rho^2+ i \epsilon}
    +\frac{1}{S^2-\rho^2 - i \epsilon}\bigg).
\eqa
To derive \eqref{eq:calD} we have systematically identified terms related by the interchange $\rho \leftrightarrow S$. This is possible because $\int_4$ is invariant under shifts and rotations of the spatial components of the vectors $\pi^\alpha_{1,2,3}$ and $\omega^\alpha$.
The $\phi$ dependence is entirely contained in ${S^2-\rho^2}$ and can be integrated out by using
\bqa
\label{eq:intphi}
&&\int_{-1}^{1} d c_\phi \Big(1-c^2_\phi\Big)^{-1/2}
    {\rm P}\bigg[\frac{1}{S^2-\rho^2}\bigg]= \nl
&&~\pi\,{\rm sgn(U)} \frac{\rm \TH(U^2-V^2)}{\sqrt{U^2-V^2}}.
\eqa
Finally, inserting \eqref{eq:intphi} and \eqref{eq:calD} into \eqref{eq:DF} gives
\bqa
\label{eq:dsub}
{\cal D}_{\rm F}(x)  = 
\int_{-\infty}^{\infty} \prod_{j=1}^2 \left( \frac{d \sigma_j}{\sigma_j-i \epsilon}\right) F_{\cal D}(\sigma_1,\sigma_2,x),
\eqa
with $F_{\cal D}(\sigma_1,\sigma_2,x)$ provided in \ref{app:b}.

Our MC estimates are presented in Table \ref{tab:1} and compared
to the analytic result \cite{Ellis:2007qk}
\bqa
\label{eq:dana}
\left. {\cal D}_{\rm F}(x) \right|_{\rm Analytic}= \frac{i \pi^2}{x} \left(\pi^2+2 i \pi \ln(x) \right).
\eqa
The time to generate $10^{6}$ MC shots on a single 2.2 GHz processor is of about 0.4\,s.

\begin{table}
\caption{
  Numerical estimates of ${\cal D}_{\rm F}(x)/(10 \,i \pi^2)$ in \eqref{eq:dsub} for several values of $x = -t/s $. The analytic result is reported in \eqref{eq:dana}.
  Numbers obtained with $4 \times 10^9 $ MC shots. MC errors between parentheses.}
\label{tab:1}       
\begin{tabular}{rll}  
\hline\noalign{\smallskip}
 $x$   & MC result & Analytic result \\
\noalign{\smallskip}\hline\noalign{\smallskip}
    .1  & 9.88(2)  $-$$i$ 1.447(1) $\times 10^{1} $  & 9.870  $-$$i$ 1.447$\times 10^{1}$\\ 
    .2  & 4.92(1)  $-$$i$ 5.055(2)                  & 4.935  $-$$i$ 5.056\\  
    .3  & 3.296(7) $-$$i$ 2.521(1)                  & 3.290  $-$$i$ 2.522\\  
    .4  & 2.476(6) $-$$i$ 1.440(1)                  & 2.467  $-$$i$ 1.439 \\
    .5  & 1.976(4) $-$$i$ 8.714(8) $\times 10^{-1}$  & 1.974  $-$$i$ 8.710$\times 10^{-1}$\\  
    .6  & 1.643(4) $-$$i$ 5.350(7) $\times 10^{-1}$  & 1.645  $-$$i$ 5.349$\times 10^{-1}$\\  
    .7  & 1.408(4) $-$$i$ 3.202(6) $\times 10^{-1}$  & 1.410  $-$$i$ 3.202$\times 10^{-1}$\\  
    .8  & 1.238(4) $-$$i$ 1.74(1)  $\times 10^{-1}$  & 1.234  $-$$i$ 1.753$\times 10^{-1}$\\
    .9  & 1.097(4) $-$$i$ 7.5(1)   $\times 10^{-2}$  & 1.097  $-$$i$ 7.356$\times 10^{-2}$  \\
\noalign{\smallskip}\hline
\end{tabular}
\end{table}

\section{A two-loop example}
\label{sec:sec7}
\begin{figure}
\vskip -4.4cm
\hskip -4.3cm
\includegraphics[width=6.5in]{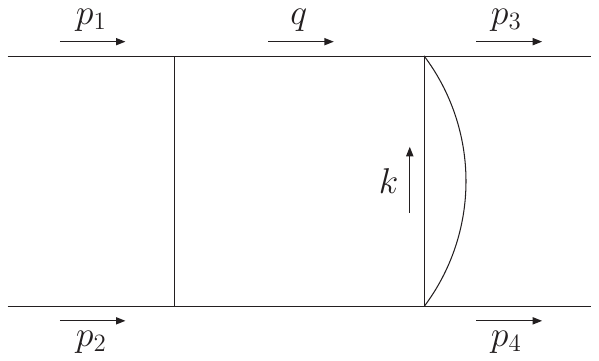}
\vskip -4.35cm
\caption{The two-loop bubble-box of \eqref{eq:E}.}
\label{fig:4}   
\end{figure}

In this section we compute the two-loop bubble-box diagram of Fig.~\ref{fig:4}, which has collinear and soft IR divergences in addition to a UV-divergent sub-diagram,
\bqa
\label{eq:E}
&&T= \mu^{8-2n}\int \frac{d^nq}{i \pi^{\frac{n}{2}}}\frac{d^nk}{i \pi^{\frac{n}{2}}} \frac{1}{D_0 D_1 D_2 D_3 D_4}, \\ 
&&
\begin{tabular}{ll}
$D_0 = q^2+i\epsilon$, &  $D_1=(q-p_1-p_2)^2+i\epsilon$, \\
$D_2=(q-p_1)^2+i\epsilon$, & $D_3= k^2+i\epsilon$, \\ 
$D_4=(k+q-p_3)+i\epsilon$, &   $p^2_{1,2,3,4}= 0$.    \nonumber
\end{tabular}  
\eqa
More precisely, we evaluate its IR and UV subtracted counterpart obtained as described and reported in \cite{Anastasiou:2018rib},
\bqa
\label{eq:Esub}
&&T_{\rm F}= \int \frac{d^4q}{i \pi^2}\frac{d^4k}{i \pi^2} 
\bigg\{\frac{1}{D_0 D_1 D_2}\bigg(\frac{1}{D_3D_4}-
\Big[\frac{1}{D_3D_4}\Big]_{q= p_1}\bigg) \nl
&&~-\bigg(\frac{1}{D_0D_2}-\frac{1}{(D_0-m^2) (D_2-m^2)}\bigg) \frac{1}{s(1-x_1)} \nl
&&~\times
\bigg(
\Big[\frac{1}{D_3D_4} \Big]_{q= x_1 p_1}
-\Big[\frac{1}{D_3D_4} \Big]_{q= p_1}
\bigg) \nl
&&~-\bigg(\frac{1}{D_1D_2}-\frac{1}{(D_1-m^2) (D_2-m^2)}\bigg) \frac{1}{s(1-x_2)} \nl
&&~\times
\bigg(
\Big[\frac{1}{D_3D_4} \Big]_{q= p_1+p_2(1-x_2)}
-\Big[\frac{1}{D_3D_4} \Big]_{q= p_1}
\bigg) \bigg\}.
\eqa
The first term of \eqref{eq:Esub} is the original integral, while
$x_1 = \frac{q \cdot p_2}{p_1 \cdot p_2}$ and  
$x_2= \frac{(p_1+p_2-q)\cdot p_1}{p_1 \cdot p_2}$ are the fractions of the momenta $p_1$ and $p_2$ carried by the internal lines with momenta $q$ and
$p_1+p_2-q$, respectively. Finally, $m$ is an arbitrary mass used to subtract the UV behavior.

The best way to apply the algorithm of Sect.~\ref{sec:sec2} to the case at hand is to use the gluing procedure described in \cite{Pittau:2021jbs}, which allows one to express \eqref{eq:Esub} in terms of a tree-level part convoluted with the one-loop sub-diagram.
To achieve this, we perform analytically the $k$ integration and rescale all dimensionful quantities by $\sqrt{s}$, which produces 
\bqa
\label{eq:Esubscal}
&&{\cal T}_F(x)\equiv s T_F= -\frac{1}{i \pi^2} \int d^4 \omega
\bigg\{\frac{1}{d_0 d_1 d_2}\,\ln\frac{d_5}{-x+i \epsilon} \nl
&&-\bigg(\frac{1}{d_0 d_2}-\frac{1}{(d_0-\mu_0) (d_2-\mu_0)}\bigg)
\frac{1}{1-x_1}  \ln\frac{x_1x-i \epsilon}{x-i\epsilon}\nl
&&-\bigg(\frac{1}{d_1 d_2}-\frac{1}{(d_1-\mu_0) (d_2-\mu_0)}\bigg)
\frac{1}{1-x_2} \ln\frac{x_2x-i \epsilon}{x-i\epsilon}
\bigg\}, \nl
\eqa
where $\mu_0= m^2/s$ and
\bqa
\label{eq:s012}
d_0 &=& \tau^2-\rho^2+i\epsilon \equiv \sigma_0+i \epsilon \nl
d_1 &=& (\tau-1)^2-\rho^2+i\epsilon \equiv \sigma_1+i \epsilon \nl
d_2 &=& \tau^2-\tau-\rho^2+\rho c_\theta+i\epsilon \equiv \sigma_2+i \epsilon \nl
d_5 &=& A+Bc_\phi +i \epsilon.
\eqa
In \eqref{eq:s012} $d_{0,1,2}= D_{0,1,2}/s$ are rescaled denominators, $d_5= [(q-p_3)^2+i \epsilon]/s$ and
\bqa
A &=& \tau^2-\tau-\rho^2+\rho c_\theta(1-2x), \nl
B &=& 2\rho s_\theta \sqrt{x(1-x)}.
\eqa
Next, we trade the integrations over $\tau$, $\rho$ and $c_\theta$ for 
integrations over $\sigma_{0,1,2}$ defined in \eqref{eq:s012}. This gives
\bqa
\int d^4 \omega=  \frac{1}{4} \int d \sigma_0 d \sigma_1 d \sigma_2\,
K(\sigma_0,\sigma_1,\sigma_2)\int_0^{2 \pi} d \phi,
\eqa
with
\bqa
\label{eq:Kdef}
K(\sigma_0,\sigma_1,\sigma_2)&=& \TH[\lambda(1,\sigma_0,\sigma_1)] \\
&&\!\!\!\!\times \TH[\sigma_0+\sigma_1-1-2\sigma_2+\lambda^{1/2}(1,\sigma_0,\sigma_1)] \nl
&&\!\!\!\!\times \TH[2\sigma_2-\sigma_0-\sigma_1+1+\lambda^{1/2}(1,\sigma_0,\sigma_1)], \nonumber
\eqa
where $\lambda(x,y,z)$ is the K\"all\'en function.
The integration over the azimuthal angle $\phi$ can be performed by using
\bqa
\label{eq:intphiT}
&&\int_0^{2 \pi}\!d \phi\, \ln(A+B c_\phi + i \epsilon) \equiv 2 \pi\,G(A,B) = \nl
&&~~~2 \pi \left[ \ln\frac{A+i\epsilon}{2}+\ln\bigg(1+\sqrt{1-\frac{B^2}{(A+i\epsilon)^2}}\bigg) \right].
\eqa
After that, ${\cal T}_{\rm F}(x)$ can be written in terms of a triple integral suitable to be evaluated numerically with {\tt GLoop},
\bqa
\label{eq:Esubfin}
{\cal T}_{\rm F}(x)  = 
\int_{-\infty}^{\infty} \prod_{j=0}^2 \left( \frac{d \sigma_j}{\sigma_j+i \epsilon}\right) F_{\cal T}(\sigma_0,\sigma_1,\sigma_2,x),
\eqa
with $F_{\cal T}(\sigma_0,\sigma_1,\sigma_2,x)$ given in \ref{app:c}. Note that the numerator $F_{\cal T}$ contains branch cuts controlled by the $i \epsilon$ prescription (see \eqref{eq:intphiT}), but because our algorithm maintains $\sigma_{0,1,2}$ in the real axis, the correct Riemann sheet is automatically taken.  

The analytic result in the physical region $0 \le x \le 1$ reads \cite{Anastasiou:2018rib} 
\bqa
\label{eq:eana}
      {\cal T}_{\rm F}(x)&=& -S_{12}(1-x) -3 \zeta_3
      -\frac{\pi^2}{3} \ln \mu_0
      + \frac{1}{6} \ln^3 x \nl
      &&+i \pi \left[{\rm Li}_2(1-x)-\frac{\pi^2}{6}
        +\frac{1}{2} \ln^2 x \right],
\eqa
where $S_{12}$ is the Nielsen polylogarithm.
Table \ref{tab:2} shows a comparison between our MC estimate based on \eqref{eq:Esubfin} and \eqref{eq:eana}. 
The time to produce $10^{6}$ MC shots on a single 2.2 GHz processor is of about 0.42\,s.

\begin{table}
\caption{Numerical estimates of ${\cal T}_{\rm F}(x)$ in \eqref{eq:Esubfin} for several values of $x = -t/s$ and $\mu_0=1$. The analytic result is reported in \eqref{eq:eana}. Numbers obtained with $8 \times 10^{10} $ MC shots. MC errors between parentheses.}
\label{tab:2}       
\begin{tabular}{rll}  
\hline\noalign{\smallskip}
 $x$   & MC result & Analytic result \\
\noalign{\smallskip}\hline\noalign{\smallskip}
    .1  &$-$ 6.19(3)   +$i$ 7.26(2)         &$-$ 6.226     +$i$ 7.244 \\ 
    .2  &$-$ 4.64(3)   +$i$ 2.29(2)         &$-$ 4.669     +$i$ 2.278\\  
    .3  &$-$ 4.11(3) $-$$i$ 8(2) $\times 10^{-2}$ &$-$ 4.136 $-$$i$ 9.670$\times 10^{-2}$\\  
    .4  &$-$ 3.86(3) $-$$i$ 1.55(2)         &$-$ 3.887   $-$$i$ 1.563 \\
    .5  &$-$ 3.73(3)  $-$$i$ 2.58(2)        &$-$ 3.756   $-$$i$ 2.584 \\  
    .6  &$-$ 3.66(3) $-$$i$  3.34(2)        &$-$ 3.683   $-$$i$ 3.346\\  
    .7  &$-$ 3.62(3) $-$$i$  3.93(2)        &$-$ 3.642   $-$$i$ 3.943\\  
    .8  &$-$ 3.60(3) $-$$i$  4.41(2)        &$-$ 3.620   $-$$i$ 4.427\\
    .9  &$-$ 3.59(3) $-$$i$  4.81(2)        &$-$ 3.609   $-$$i$ 4.828\\
\noalign{\smallskip}\hline
\end{tabular}
\end{table}

\section{Conclusion and outlook}
\label{sec:sec8}
Any attempts towards a numerical loop integration requires controlling threshold singularities. A possible approach is contour deformation \cite{Soper:1999xk,Capatti:2019edf}, that calls for analytic knowledge of the cut structure of the integrand \cite{Kermanschah:2021wbk} or numerical checks establishing whether the deformation stays on the correct side of the singularity.
In \cite{Pittau:2021jbs} an alternative has been proposed and shown to be effective in the MC estimate of four-dimensional multi-loop integrals directly in Minkowski space.

In this paper we have extended this technique to massless scalar one-loop integrals with no more that four external legs, regularized within dimensional regularization.
Our strategy is based on a separation of the $1/\varepsilon$ and $1/\varepsilon^2$ poles before integration. The finite part, containing the threshold singularities, can then be integrated numerically in four dimensions.
A fully numerical evaluation of one-loop amplitudes with more than four legs is in principle possible if the coefficients of the contributing one-, two-, three- and four-point integrals are also determined in a numerical fashion by using, for instance, the method of \cite{Ossola:2006us}.

We have presented numerical results obtained with the help of the code {\tt GLoop}.
A MC error of the order of a few per mil can usually be obtained for a modest CPU cost.
As with any other numerical method, this level of precision is expected to be sufficient for phenomenological purposes when the gauge cancellations are moderate. If this is not the case, cancellations among diagrams must be enforced to occur before integration. This should be feasible because they are usually controlled by Ward identities operating at the integrand level.

Enlarging the range of applicability of our method beyond one loop requires removing UV and IR divergences by adding appropriate counterterms at the level of the integrand. Ultraviolet counterterms can be constructed by using, for instance, the same procedure that defines FDR integrals \cite{Pittau:2012zd,Donati:2013iya,Pittau:2013qla,Donati:2013voa,Page:2015zca,Page:2018ljf}. As for the infrared behavior, a systematic approach has been developed for two-loop integrals by Anastasiou and Sterman in \cite{Anastasiou:2018rib}.
In Sect.~\ref{sec:sec7} we have presented a simple two-loop example showing how the subtraction method of \cite{Anastasiou:2018rib} can be combined with our algorithm. 
For all these reasons, we believe that the strategy described in this paper can be extended up to two loops. A deeper exploration of this issue is planned for a future publication.

\appendix
\section{$F_{\cal C}(\sigma_1,\sigma_2)$}
\label{app:a}
The numerator of the subtracted 3-point function of \eqref{eq:csub1} reads
\bqa
&& F_{\cal C}(\sigma_1,\sigma_2) = -2 i \pi^2 \TH(\sigma_1) 
  \Big\{\TH(\sigma_3)\Big[\TH(1-\sigma_-)\TH(\sigma_-) \nl
    &&~-\TH(1+\sigma_-)\TH(-\sigma_-)-\TH(1-\sigma_4) \TH(\sigma_4) \nl
    &&~+\TH(1+\sigma_4)\TH(-\sigma_4)
    \Big]+\TH(-\sigma_3)\Big[\TH(\sigma_-)\TH(\sigma_+) \nl
    &&~-\TH(\sigma_-+f)\TH(\sigma_+-f)-\TH(1+\sigma_-)\TH(\sigma_+-1)
  \nl
  &&~+\TH(1+\sigma_4)\TH(\sigma_2-1)-\TH(\sigma_-+g)\TH(\sigma_+-g) \nl
 &&~+\TH(\sigma_5-1) \TH(1-\sigma_2)\Big] \Big\}, 
\eqa
where we have defined
\bqa
\label{eq:defsigma}
\!\!\begin{tabular}{lll}
$g= \frac{1+\sqrt{1-4 \sigma^2_1}}{2}$, & $f= \frac{\sigma^2_1}{g}$, &
$\sigma_3 = \sigma_1-\frac{1}{2}$, \\
$\sigma_4= 2 \sigma_1-\sigma_2$,& $\sigma_5= 2 \sigma_1+\sigma_2$, & $\sigma_{\pm}= \sigma_1 \pm \sigma_2$.  
\end{tabular}
\eqa

\section{$F_{\cal D}(\sigma_1,\sigma_2,x)$}
\label{app:b}
The numerator of the subtracted 4-point function of \eqref{eq:dsub} is
\bqa
\label{eq:FD}
&& F_{\cal D}(\sigma_1,\sigma_2,x) = 4 i \pi^2 \TH(\sigma_1) \nl 
&&\times \big\{
     \TH(\sigma_-) \TH(U_1-|\sigma_3|) \big [N_1-1/x\big] \nl
&&~\!-\TH(\sigma_4) \TH(U_2-|\sigma_3|) \big[(1-2\sigma_1)N_2-1/x \big] \nl
&&~\!-\TH(1+\sigma_-) \TH(U_3-|\sigma_3|) \big[N_3-1/x \big] \nl
     &&~\!+\TH(1+\sigma_4) \TH(U_4-|\sigma_3|) \big[(1+2\sigma_1)N_4-1/x \big] \big\} \nl
&&-i\frac{\pi^2-\ln^2(x)}{x}\TH(1-|\sigma_1|)\TH(1-|\sigma_2|), 
\eqa
where  we have used the definitions in \eqref{eq:defsigma} and
\bqa
\begin{tabular}{ll}
$U_1= \sigma_+-\sigma_3$, & $U_2= \sigma_2-\sigma_3$, \nl
$U_3= \sigma_3-\sigma_-$, & $U_4= \sigma_3-\sigma_4$. 
\end{tabular}
\eqa
Furthermore
\bqa
N_i= {\rm sgn}\big[(1+2\sigma_1)(1-2\sigma_1)+4 U^2_i\big]\frac{\TH(W_i)}{\sqrt{W_i}},
\eqa
with
\bqa
W_i= \frac{x}{4} \bigg\{
16 xU^2_i+\!\!
\prod_{\lambda_1, \lambda_2= \pm}
\big[1+2(\lambda_1\sigma_1+\lambda_2 U_i)\big]\bigg\}. 
\eqa
Note that the last term of \eqref{eq:FD} generates the unintegrated contribution of \eqref{eq:rescdsub}. 

\section{$F_{\cal T}(\sigma_0,\sigma_1,\sigma_2,x)$}
\label{app:c}
The numerator of the subtracted two-loop bubble-box of \eqref{eq:Esubfin} reads
\bqa
&&F_{\cal T}(\sigma_0,\sigma_1,\sigma_2,x) = \nl
&&~~\frac{1}{2 i \pi} \bigg\{ K(\sigma_0,\sigma_1,\sigma_2) \bigg[\ln(-x+i \epsilon)-G(A,B)\nl
&&~~+ \frac{\sigma_1}{\sigma_1-\sigma_2}\ln \frac{x(1+\sigma_2-\sigma_1)-i \epsilon}{x-i\epsilon} \nl
&&~~+ \frac{\sigma_0}{\sigma_0-\sigma_2}\ln \frac{x(1+\sigma_2-\sigma_0)-i \epsilon}{x-i\epsilon} \bigg] \\
&&~~-\frac{\sigma_1}{\sigma_1-\bar \sigma_2}\ln \frac{x(1+\bar \sigma_2-\sigma_1)-i \epsilon}{x-i\epsilon}\,
K(\bar \sigma_0,\sigma_1,\bar \sigma_2) \nl
&&~~- \frac{\sigma_0}{\sigma_0-\bar \sigma_2}\ln \frac{x(1+\bar \sigma_2-\sigma_0)-i \epsilon}{x-i\epsilon}\,
K(\sigma_0,\bar \sigma_1,\bar \sigma_2) \bigg\}, \nonumber
\eqa
where $\bar \sigma_i = \sigma_i+\mu_0$.
The functions $K(\sigma_0,\sigma_1,\sigma_2)$ and  $G(A,B)$ are defined in \eqref{eq:Kdef} and \eqref{eq:intphiT}, respectively.
Note that in terms of $\sigma_{0,1,2}$ one has
\bqa
1-x_1 &=& \sigma_1-\sigma_2, \nl
1-x_2 &=& \sigma_0-\sigma_2
\eqa
and
\bqa
A &=& x(\sigma_0+\sigma_1-1) +\sigma_2 (1-2 x), \nl
B &=& 2 \sqrt{x(1-x)}\sqrt{(\sigma_2-\sigma_1)(\sigma_0-\sigma_2)-\sigma_2}.
\eqa

\acknowledgement
I would like to thank Bryan Webber for constructive criticism of the manuscript.

\noindent This work has been partially funded by

\noindent MICIU/AEI/10.13039/501100011033 and ERDF/EU (grant PID2022-139466NB-C22) and by the SRA grant PID2019-106087GB-C21 (10.13039/501100011033).

\bibliographystyle{spphys}       
\bibliography{mcloopsi}   

\end{document}